\documentclass[%
 reprint,
 superscriptaddress,
 amsmath,amssymb,
 aps,longbibliography
]{revtex4-1}

\usepackage{graphicx}
\usepackage{dcolumn}
\usepackage{bm}
\usepackage{hyperref}
\usepackage[mathlines]{lineno}
\usepackage[bottom]{footmisc}
\usepackage[super]{nth}%
\usepackage{color}
\usepackage{hhline}
\usepackage{physics}
\usepackage{amsmath}
\usepackage{lipsum}

\begin{document}
\raggedbottom

\title{Aperiodic bandgap structures for enhanced quantum two-photon sources}
\author{Luca Dal Negro}
\email{dalnegro@bu.edu}
\affiliation{Department of Electrical and Computer Engineering, Boston University, 8 Saint Mary\textsc{\char13}s Street, Boston, Massachusetts 02215, USA}
\affiliation{Division of Material Science and Engineering, Boston University, 15 Saint Mary\textsc{\char13}s Street, Brookline, Massachusetts 02446, USA}
\affiliation{Department of Physics, Boston University, 590 Commonwealth Avenue, Boston, Massachusetts 02215, USA}
\author{Yuyao Chen}
\affiliation{Department of Electrical and Computer Engineering, Boston University, 8 Saint Mary\textsc{\char13}s Street, Boston, Massachusetts 02215, USA}
\author{Sean Gorsky}
\affiliation{Department of Electrical and Computer Engineering, Boston University, 8 Saint Mary\textsc{\char13}s Street, Boston, Massachusetts 02215, USA}
\author{Fabrizio Sgrignuoli}
\affiliation{Department of Electrical and Computer Engineering, Boston University, 8 Saint Mary\textsc{\char13}s Street, Boston, Massachusetts 02215, USA}

\begin{abstract}
In this paper we propose a novel approach to enhance the efficiency of the two-photon spontaneous emission process that is driven by the multifractal optical mode density of photonic structures based on the aperiodic distributions of Eisenstein and Gaussian primes.
In particular, using the accurate Mie-Lorenz multipolar theory in combination with multifractal detrended fluctuation analysis, we compute the local density of states of periodic and aperiodic systems and demonstrate the formation of complete bandgaps with distinctive fractal scaling behavior for scattering arrays of dielectric nanocylinders. Moreover, we systematically study the Purcell enhancement and the most localized optical mode resonances in these novel aperiodic photonic systems and compute their two-photon spontaneous emission rates based on the general Green's tensor approach. Our results demonstrate that the excitation of the highly-resonant critical states of Eisenstein and Gaussian photonic arrays across broadband multifractal spectra gives rise to significantly enhanced emission rates compared to what is possible at the band-edges of periodic structures with comparable size. Besides defining a novel approach for enhanced quantum two-photon sources on the chip, the engineering of aperiodic bandgap structures with multifractal mode density may provide access to novel electromagnetic resonant phenomena in a multiscale-invariant vacuum for quantum nanophotonics applications.  
\end{abstract}

\maketitle
\section{Introduction}
The generation and manipulation of correlated multiphoton states play a central role in quantum communication technology and cryptography \cite{Obrien,Metcalf,Wang,Grosshans,RevModPhys.74.145}. In particular, quantum states that involve entangled photon pairs have found numerous applications to biological imaging, quantum sensing and computing where they are typically generated on solid state platforms using quantum dots transitions and nonlinear processes such as spontaneous parametric down-conversion (SPDC) \cite{RevModPhys.84.777,RevModPhys.89.035002,Liu19,orieux2017semiconductor}. 
Recently, the demonstration of  two-photon spontaneous emission (TPSE) at room-temperature in semiconductor structures has attracted a significant attention as an alternative approach for the generation of entangled photon pairs \cite{HayatNaturePhot,Driel,Hayat,Ota}. In the TPSE quantum process, an excited electron decays to its ground state by simultaneously emitting a pair of photons that are time-energy entangled \cite{Hayat}. Differently from one-photon emission processes, any combination of photon energies satisfying the constraint of total energy conservation is allowed in the TPSE process, which results in very broad emission spectra. In TPSE processes, emitters with discrete energy levels behave as radiation sources with continuous spectra. However, since TPSE is a second-order process in perturbation theory, the decay rate in bulk materials is generally several orders of magnitude (5 to 8) smaller than the one of competing one-photon emission channels, limiting the applicability to quantum radiation sources \cite{Craig}. While inherently much weaker compared to single photon processes, TPSE has nevertheless played an important role in astrophysics, where it explained the broadband emission of planetary nebulae, and it has been deeply investigated in atomic physics starting from the pioneering theoretical paper by G\"{o}ppert-Mayer \cite{Goppert} and the numerical analysis of the TPSE rates in hydrogen and helium by Breit and Teller \cite{Breit}.

Recent breakthroughs in nano-optical materials and technologies have attracted renewed interest in the engineering of TPSE quantum phenomena at the nanoscale \cite{Poddubny,MunizPRL}.    
In particular, by tailoring the local density of photon states (LDOS) in nanostructured environments \cite{LodahlReview}, dramatic enhancements in the TPSE emission rate have been demonstrated using photonic cavities \cite{Ota,Li}, plasmonic and two-dimensional nanostructures \cite{Gonccalves,Poddubny,Nevet}, and phonon-polaritons dielectric structures \cite{Rivera}. More broadly, LDOS engineering in plasmonic and polaritonic nanostructures with sub-wavelength field confinement has been recently established for tailoring and enhancing TPSE rates to levels that can even exceed the rates of spontaneous one-photon processes by nearly two orders of magnitude \cite{Rivera}. Thanks to the unique ability to shape the emission spectra of photon pairs by controlling the resonant frequencies of the medium, TPSE engineering is rapidly emerging as a  promising new approach for high-rate multiphoton sources on the chip. Deterministic aperiodic systems are artificial optical media with a tunable degree of structural complexity that ranges in between periodic crystals and disordered structures \cite{DalNegroReview}. They are generated by deterministic mathematical rules and  have emerged as an alternative material platform for designing novel complex photonic devices for both classical and quantum photonics applications \cite{DalNegroReview,Trojak1,Trojak2,Trevino,TrevinoNano,Macia}.

In this paper, using the rigorous two-dimensional generalized Lorenz-Mie theory (2d-GMT) \cite{Gagnon,AsatryanPRE}, we investigate the LDOS and TPSE enhancement spectra in finite-size scattering arrays of dielectric nanocylinders arranged according to the aperiodic distributions of the prime elements in the Eisenstein and Gaussian integer rings \cite{lang2013algebraic}. These deterministic aperiodic photonic structures, simply referred to as Eisenstein and Gaussian arrays, possess fascinating structural, transport, and spectral properties \cite{WangPRB,SgrignuoliRW} and support characteristic scattering resonances with strong multifractal behavior akin to localized modes near the Anderson transition of random systems \cite{SgrignuoliMF}. 
\begin{figure*}[t!]
\centering
\includegraphics[width=\linewidth]{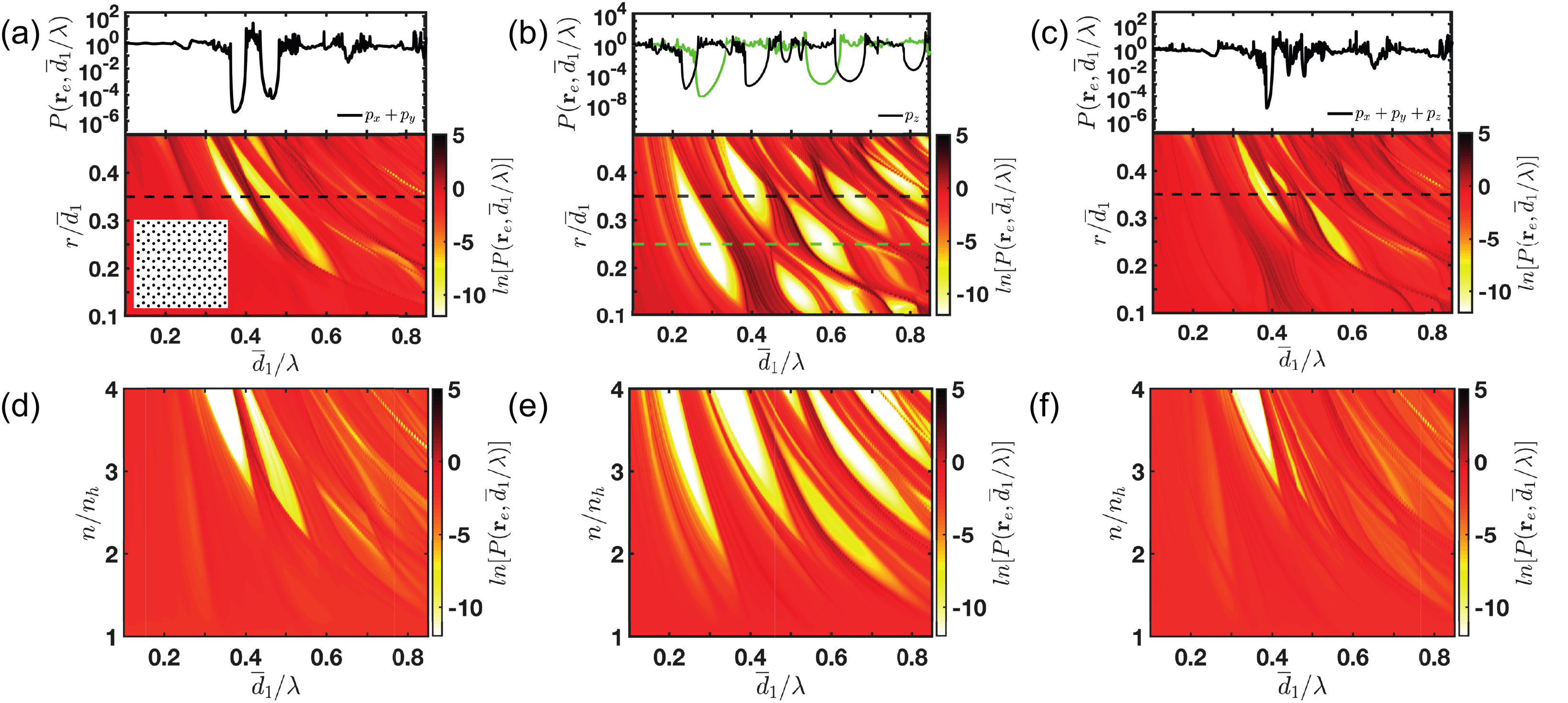}
\caption{Purcell enhancement maps of 298 dielectric nanocylinders ($\varepsilon=10.5$) arranged in a honeycomb lattice [as shown in the inset of panel (a)] as a function of $\overline{d}_1/\lambda$ and the ratio between particle radius $r$ and averaged center-to-center distance $\overline{d}_1$. The 2D-maps of panels (a), (b), and (c) are evaluated by probing the structures with two horizontal (i.e, along the plane of the nanocylinders) line sources, one vertical (i.e, perpendicular to the plane of the nanocylinders) line source, and by a linear combination of them located at the center (i.e., $\textbf{r}_e=(0,0)$). All these maps are color-coded according to the $\ln[P(\textbf{r}_e,\overline{d}_1/\lambda)]$. Insets on the top of each panel show representative high-resolution (i.e., 1.5$\times10^{-4}\overline{d}_1/\lambda$ spectral resolution) 1D-cuts of the $P(\textbf{r}_e,\overline{d}_1/\lambda)$ plotted in semi-logarithmic scale. Panels (d), (e), and (f) show the Purcell enhancement maps  corresponding to the atlas gap maps shown directly above and evaluated  by sweeping $\overline{d}_1/\lambda$ and the refractive index contrast $n/n_h$.}
\label{Fig1}
\end{figure*}
By computing the electromagnetic Green tensor, we investigate the LDOS for both transverse and vertical polarizations, showing the formation of complete photonic bandgaps with a characteristic fractal scaling behavior. Our results are systematically compared to the case of periodic arrays of nanocylinders arranged in the honeycomb lattice, which are structures with a complete photonic bandgap but without any fractal features \cite{Joannopoulos}. This comparison will allow us to underline the main differences between the standard and fractal gaps. Furthermore, using multifractal detrended fluctuation analysis (MDFA) \cite{Kantelhardt}, we establish the multifractal nature of the LDOS spectra of the Eisenstein and Gaussian arrays. This feature reflects the multifractal properties of the geometrical support of the investigated structures. Multifractals are inhomogeneous complex systems with multi-scale structural fluctuations that manifest a characteristic type of long-range correlations described by local power-law scaling laws \cite{falconer2004fractal,SgrignuoliRW}.
Finally, we compute the TPSE enhancement spectra within the Green's function spectral method \cite{MunizPRA, MunizPRL} and show how the presence of broadband LDOS fluctuations with multifractal scaling significantly enhance the TPSE rates compared to traditional photonic crystals with smooth (non-fractal) mode density. The engineering of complex scattering media with multifractal LDOS spectra provides a new route to achieve broadly tunable sources of entangled photon pairs based on room-temperature TPSE for quantum information research. 
\section{Fractal bandgaps of prime arrays}
\begin{figure*}[t!]
\centering
\includegraphics[width=\linewidth]{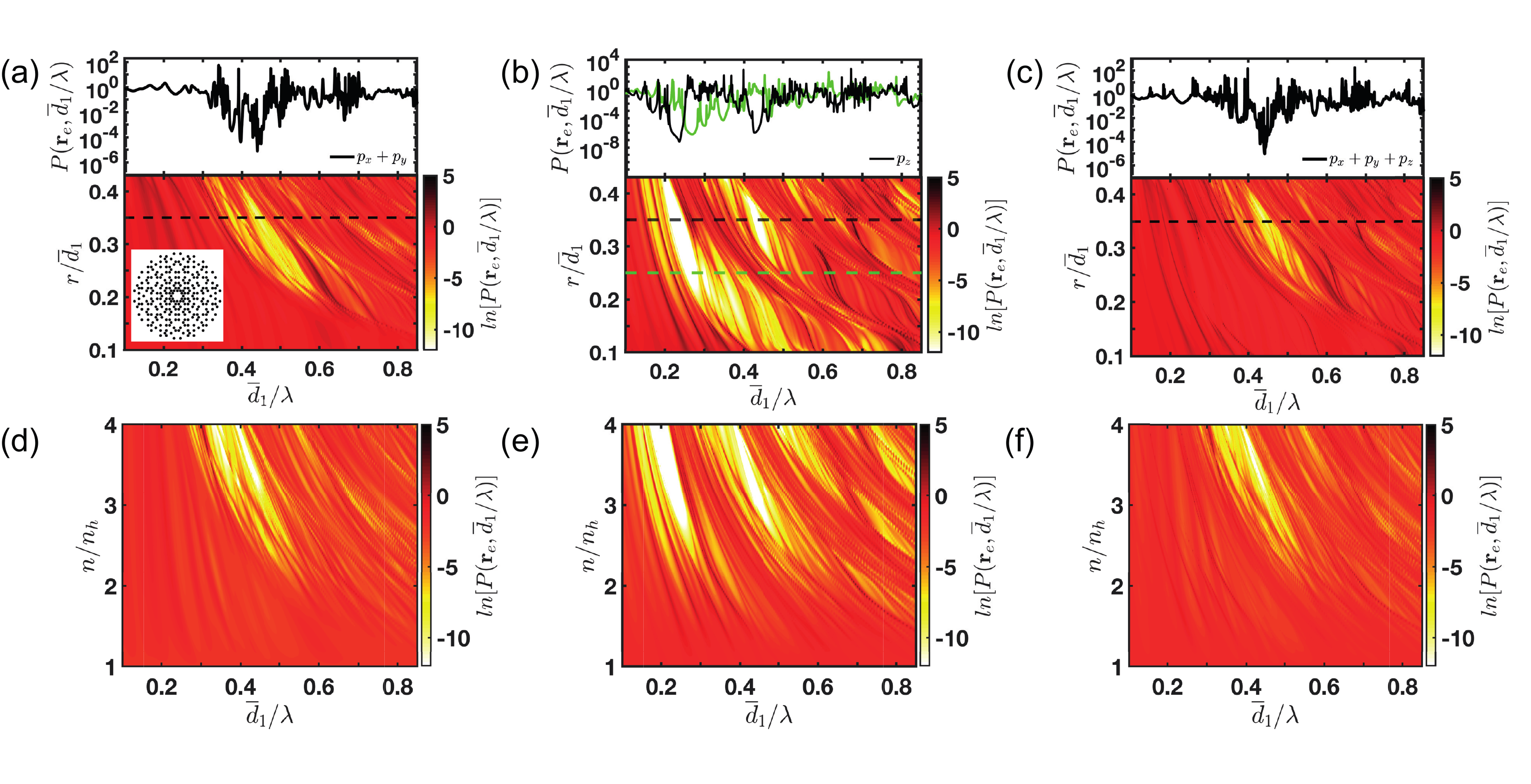}
\caption{Purcell enhancement maps of 276 dielectric nanocylinders ($\varepsilon=10.5$) arranged by following the distribution of Eisenstein primes [as shown in the inset of panel (a)] as a function of $\overline{d}_1/\lambda$ and $r/\overline{d}_1$. The 2D-maps of panels (a), (b), and (c) are evaluated by probing the structures with two horizontal line sources, one vertical line source, and by a linear combination of them located at the center. All these maps are color-coded according to the $\ln[P(\textbf{r}_e,\overline{d}_1/\lambda)]$. Insets on the top of each panels show representative high-resolution (i.e., 1.5$\times10^{-4}\overline{d}_1/\lambda$ spectral resolution) 1D-cuts of $P(\textbf{r}_e,\overline{d}_1/\lambda)]$ plotted in semi-logarithmic scale. Panels (d), (e), and (f) show the Purcell enhancement maps  corresponding to the atlas gap maps shown directly above and evaluated  by sweeping $\overline{d}_1/\lambda$ and the refractive index contrast $n/n_h$.}
\label{Fig2}
\end{figure*}
In order to investigate the dynamics of a radiation source embedded into a nanostructured environment we compute the corresponding Purcell enhancement factor, from which the spectral locations of the resonant modes with the highest quality factors, such as those that are near the bandgap regions \cite{SgrignuoliACS,John,Trevino,DalNegrocrystal2016,Trojak1,Trojak2,DalNegroReview,TrevinoNano}, can be readily identified. The Purcell factor 
describes the modifications of the spontaneous decay properties of a line source (i.e., the 2d equivalent of an emitting dipole) located at $\textbf{r}_e$ due to its coupling with the surrounding photonic environment \cite{Novotny}. Therefore, the study of the Purcell factor of a localized source in a photonic structure allows one to quantify the light emission enhancement or suppression due to the multiple light scattering effects in the considered structures. Moreover, in the absence of material absorption, the Purcell factor coincides with the enhancement of the radiative rate of the system \cite{vogel2006quantum}.

In this work we have used the rigorous theory of multipole expansions \cite{Gagnon} with modifications to allow for interior and exterior line-sources \cite{AsatryanPRE}, and we evaluated the Green's tensor $\mathbb{G}(\textbf{r},\textbf{r};\omega)$ from which the Purcell factor $P(\textbf{r}_e;\omega)$ is obtained as follows \cite{Sheng,AsatryanPRE}:
\begin{equation}\label{2DLDOS}
P(\textbf{r}_e;\omega)=\frac{\rho(\textbf{r}_e,\omega)}{\rho_0(\omega)}=-4n_0^2\Im[\mathbb{G}(\textbf{r}_e,\textbf{r}_e;\omega)]
\end{equation}
Here, $\rho(\textbf{r}_e,\omega)$ and $\rho_0(\omega)=\omega/(2\pi c^2)$ are the LDOS of the structure and the free-space LDOS, respectively, while $n_0$ is the refractive index of the embedding medium. Note that $\mathbb{G}(\textbf{r},\textbf{r}_e;\omega)$ describes the electric field response at the observation point $\textbf{r}$ due to a line source located at $\textbf{r}_e$.  
\begin{figure*}[t!]
\centering
\includegraphics[width=\linewidth]{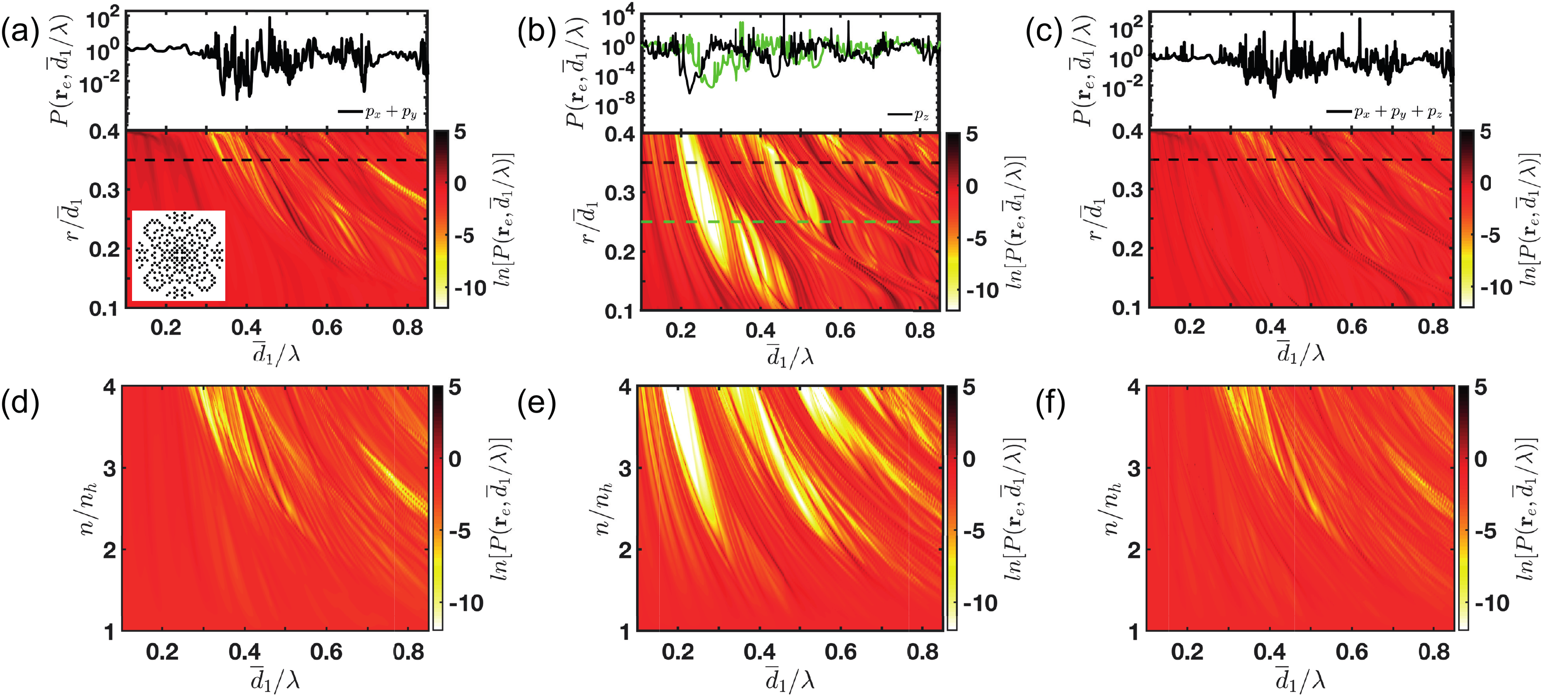}
\caption{Purcell enhancement maps of 264 dielectric nanocylinders ($\varepsilon=10.5$) arranged by following the distribution of Gaussian primes [as shown in the inset of panel (a)] as a function of $\overline{d}_1/\lambda$ and $r/\overline{d}_1$. The 2D-maps of panels (a), (b), and (c) are evaluated by probing the structures with two horizontal line sources, one vertical line source, and by a linear combination of them located at the center. All these maps are color-coded according to the $\ln[P(\textbf{r}_e,\overline{d}_1/\lambda)]$. Insets on the top of each panels show representative high-resolution (i.e., 1.5$\times10^{-4}\overline{d}_1/\lambda$ spectral resolution) 1D-cuts of the $P(\textbf{r}_e,\overline{d}_1/\lambda)$ plotted in semi-logarithmic scale. Panels (d), (e), and (f) show the Purcell enhancement maps  corresponding to the atlas gap maps shown directly above and evaluated by sweeping $\overline{d}_1/\lambda$ and the refractive index contrast $n/n_h$.}
\label{Fig3}
\end{figure*}
Since we are dealing with a 2d problem, the field solutions can be decomposed into two independent polarization bases: the transverse-magnetic (TM), where the electric field $\textbf{E}$ is oriented in the $\hat z$ direction while the magnetic field $\textbf{H}$ is in the $xy$-plane, and the transverse-electric (TE), in which the two field components are swapped. Regardless of polarization, all the relevant calculations are performed on the z-oriented field and the other components follow directly from the Maxwell's equations.
In our numerical simulations, we used an efficient algorithm based on the rigorous solution of the 2d scattering and eigenvalue problems in the framework of the generalized Mie theory \cite{AsatryanPRE}. Similarly to the case of the single particle Mie theory \cite{Bohren}, this accurate technique expands the incident, scattered, and internal electric and magnetic fields as an infinite sum of a complete set of basis functions, i.e., cylindrical Bessel functions. The polarization-dependent Mie-coefficients of the field expansions are then determined by enforcing the Maxwell's boundary conditions on the surface of each cylinder. In order to apply the boundary conditions at each cylinder, the Graf's addition theorem \cite{Abramowitz} must be used, leading to the linear system $\textbf{T}\textbf{b}=\textbf{a}_0$ with $\textbf{a}_0=a_{n\ell}/J_\ell(k_0r_n)$ and $\textbf{b}=b_{n\ell}/J_\ell(k_0r_n)$ where $k_0$ is the plane wave excitation wavenumber, $r_n$ is the radius of the $n^{th}$ nanocylinder, and coefficients $a_{n\ell}$ and $b_{n\ell}$ are the source and scattering coefficients for the $n^{th}$ nanocylinder of order $\ell$. The functions $J_\ell$ is the cylindrical Bessel function of order $\ell$. The transition matrix $\textbf{T}$ describing the material and the geometrical properties of the scattering medium is defined as \cite{Gagnon}:
\begin{equation}\label{Tmatrix}
\textbf{T}_{nn^\prime}^{\ell\ell^\prime}=\delta_{nn^\prime}\delta_{\ell\ell^\prime}-(1-\delta_{nn^\prime})e^{i(\ell^\prime-\ell)\phi_{nn^\prime}}H^{(+)}_{\ell-\ell^\prime}(k_0R_{nn^\prime})s_{n\ell}\frac{J_{\ell^\prime}(k_0r_{n^\prime})}{J_{\ell}(k_0r_{n})}
\end{equation}   
Here, $n$ indicates the $n^{th}$-cylinder and $\ell$ is the multipolar order of the cylindrical Bessel $J_\ell(x)$ and Hankel $H_\ell(x)$ functions ($x=k_0r$ is the size parameter) used in the calculations \cite{Gagnon,AsatryanPRE}. Equation\,(\ref{Tmatrix}) ensures that the matrix $\textbf{T}$ remains numerically stable during the inversion even when a large number of multipolar orders is required \cite{Gagnon}. Calculating the LDOS $\rho(\textbf{r}_e,\omega)$, and therefore its corresponding Purcell factor $P(\textbf{r}_e,\omega)$ through eq.\,(\ref{2DLDOS}), requires solving the eq.\,(\ref{Tmatrix}) with the input coefficients $\textbf{a}_0$ describing a line source and then evaluating the total fields at the position of the excitation. The electric line source is oriented along the $\boldsymbol{\hat{x}}$ and $\boldsymbol{\hat{y}}$ directions for TE polarization, and along the $\boldsymbol{\hat{z}}$ direction in the TM configuration. Finally, we remark that although our modeling approach is valid for arbitrary 2d systems, it  can also be conveniently utilized as the starting point for the design of three-dimensional photonic structures in the device-relevant membrane geometry \cite{Trojak1,Trojak2}.

In Figures\,(\ref{Fig1}-\ref{Fig3}) we show a comprehensive analysis of the Purcell enhancement computed by considering horizontally polarized line sources (in panels a,d), vertically polarized ones (in panels b,e), and a linear combination of them (in panels c,f) always located at the center of each investigated structure, i.e, $\textbf{r}_e=(0;0)$. We consider scattering systems containing approximately 300 dielectric nanocylinders of constant permittivity $\varepsilon=10.5$, embedded in air and spatially arranged according to the honeycomb lattice \cite{Joannopoulos}, the Eisenstein, and the Gaussian prime arrays \cite{SgrignuoliMF,WangPRB,SgrignuoliRW}. In the panels (a-c) of Fig.\,(\ref{Fig1}), Fig.\,(\ref{Fig2}), and Fig.\,(\ref{Fig3}) we display in logarithmic scale the Purcell factor values as a function of $\overline{d}_1/\lambda$ and $r/\overline{d}_1$ where $r$ is the particle radius and $\overline{d}_1$ the average center-to-center particle distance. On the other hand, in panels (d-f) we show the corresponding "gap recipes" obtained by calculating the Purcell enhancement maps as a function of the normalized distance $\overline{d}_1/\lambda$ and the refractive index contrast $n/n_h$. In our calculations we have considered a total of 101 values of the $r/\overline{d}_1$ and $n/n_h$ parameters and a spectral resolution $\Delta(\overline{d}_1/\lambda)=7\times10^{-4}$ for each array. These results enable the design of aperiodic bandgap structures in any desired spectral range compatible with the high transparency window of the considered dielectric materials. In fact, the spectral location of the different bandgaps can be properly adjusted by tuning the center-to-center averaged interparticle separation $\overline{d}_1$. For example, when $\overline{d}_1=220\,nm$, the fundamental TM bandgap of the investigated structures consisting of nanocylinders with a diameter of $154\,nm$ spans the spectral range $[0.73,1.1]~\mu m$ [see black dashed line in panels (b) of Figs.\,(\ref{Fig1}-\ref{Fig3})]. We note that outside of the bandgap regions, the Purcell enhancement factor fluctuates around unity, while its sharp variations correspond to the excitation of optical modes with high-quality factors in the photonic structure. Moreover, the formation of photonic bandgaps can be readily detected by the appearance of frequency regions with strongly suppressed Purcell factor in finite-size photonic structures \cite{AsatryanPRE,Trojak1,Trojak2}. 

The honeycomb lattice, shown in Fig.\,(\ref{Fig1}), displays large spectral gaps for both TE [see panel (a)] and TM [see panel (b)] polarizations. The fundamental gap region and higher-order ones are clearly visible, shifting to higher frequencies as we decrease the separation $\overline{d}_1$ between the dielectric pillars, as displayed in the inset of panel (b) for two representative $r/\overline{d}_1$ values. On the other hand, these spectral gaps red-shift by increasing the refractive index contrast, as shown in panels (d-f) by fixing the radius-pitch ratio $r/\overline{d}_1=0.35$. As it is well-known, in this lattice symmetry the fundamental TE-gap spectrally overlaps with the higher order TM-gaps, forming complete bandgaps [see panel (c)] , i.e., forbidden spectral regions for both polarizations \cite{Joannopoulos}. Interestingly, despite their lack of periodicity we discover that the Eisenstein and Gaussian structures also exhibit clear photonic bandgaps and pseudogaps (i.e., secondary gaps of smaller amplitudes generated by the local scale invariance symmetry, or multifractality, of the investigated arrays discussed in more details in the next section \cite{jiang2005photonic,ryu1992extended,dal2003light}), which are identified by the bright yellow regions in Figs.\,(\ref{Fig2}) and (\ref{Fig3}) respectively. However, in stark contrast with the situation of the honeycomb periodic structure, in the Eisenstein and Gaussian arrays we observe a characteristic fragmentation of the main gap regions into a series of smaller sub-gaps separated by a complex spectral distribution of localized modes. The splitting behavior of gap regions can also be clearly observed in the Purcell enhancement profiles displayed on top of each panel and corresponding to the $r/\overline{d}_1$ values indicated by the dashed lines of the same color. The broadband, highly irregular arrangements of gap regions and resonant states discovered in the Eisenstein and Gaussian arrays manifest their characteristic multifractal geometry \cite{SgrignuoliMF,SgrignuoliRW}. We also remark that, since the Eisenstein structure has a larger asymptotic density compared to both the honeycomb and the Gaussian ones, it supports lower frequency gaps at smaller values of the refractive index contrast [compare the panels (d-f) of Fig.\,(\ref{Fig2}) with the ones of Fig.\,(\ref{Fig1}) and Fig.\,(\ref{Fig3})]. Therefore, the Eisenstein aperiodic array is a good candidate photonic system for the engineering of polarization-insensitive bandgap structures with multifractal scaling properties, which will be addressed in the next section.

\section{Multifractal scaling of the LDOS}
\begin{figure}[t!]
\centering
\includegraphics[width=\linewidth]{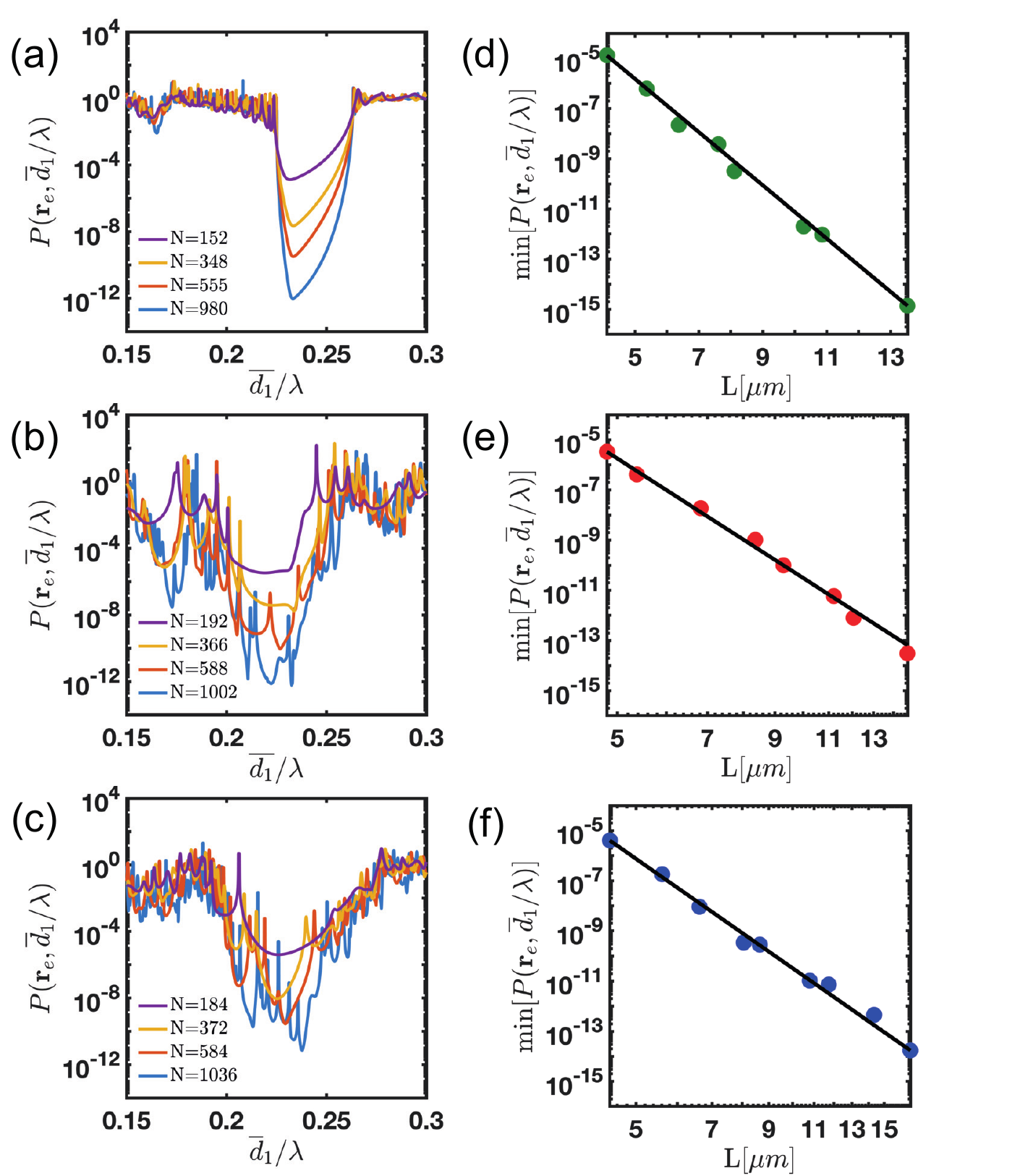}
\caption{Panels (a), (b), and (c) show the Purcell factor as a function of $\overline{d}_1/\lambda$ around the lowest energy band gap region for a different number of dielectric nanocylinders, as specified in the corresponding legends, arranged in a honeycomb lattice, Eisenstein and Gaussian configuration, respectively. These $P(\textbf{r}_e,\omega)$ profiles correspond to $r/\overline{d}_1=0.35$ and are evaluated by probing the different structures with a vertical line source located at the center. Panels (d),(e), and (f) display the scaling of the minimum value of the Purcell factor inside the bandgap as a function of the system sizes.}
\label{Fig4}
\end{figure}
\begin{figure}[b!]
\centering
\includegraphics[width=\linewidth]{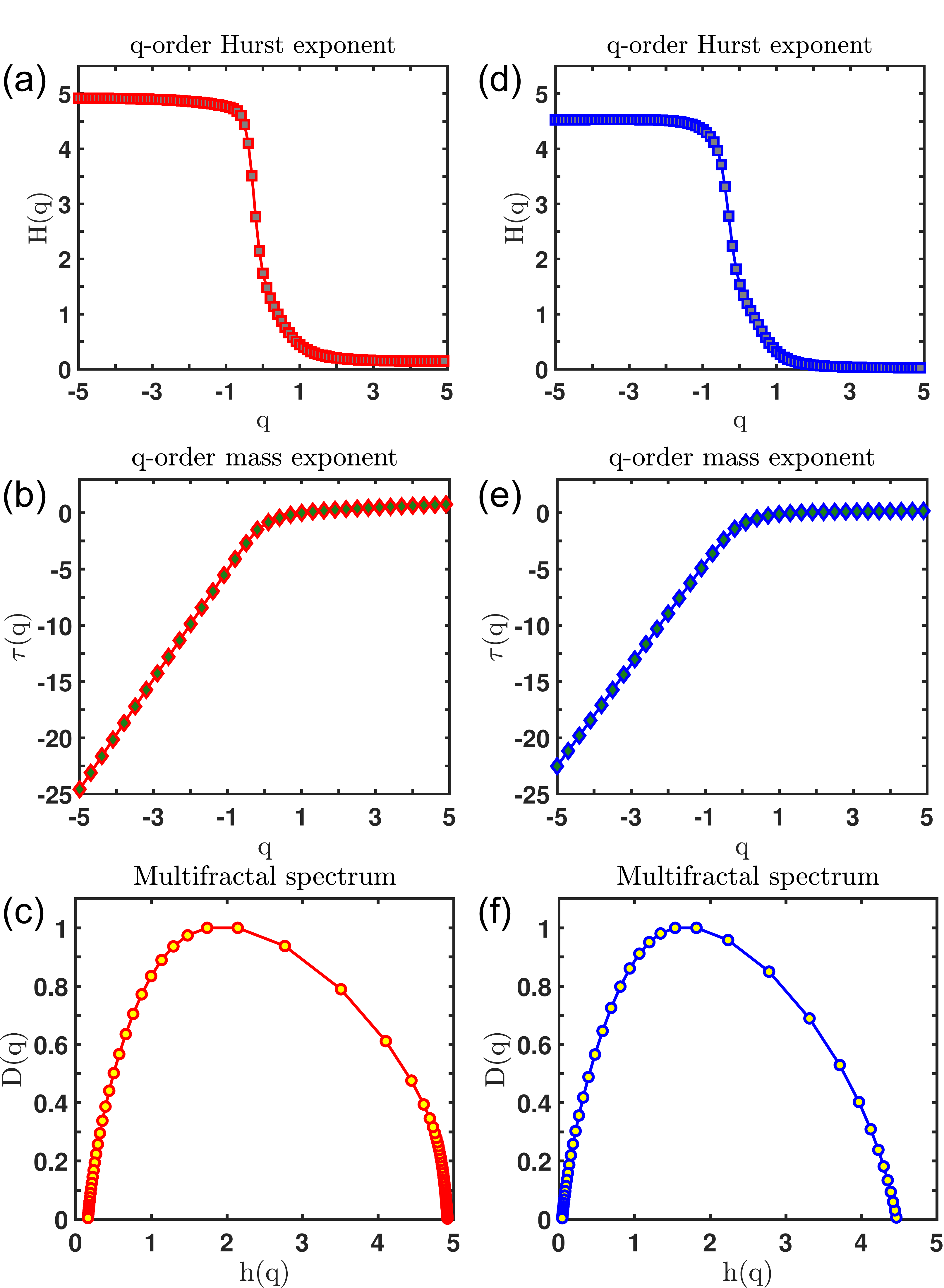}
\caption{Multifractals exponents extrapolated from the LDOS $\rho(\textbf{r}_e,\omega)$ of Eisenstein (a-c) and Gaussian array (d-f). While panel (a) and (d) show the $q$-dependence of the Hurst exponent $H(q)$, panel (b) and (e) display the  mass exponent $\tau(q)$ in the Eisenstein and Gaussian array, respectively. Panels (c) and (f) report the corresponding singularity spectra $D(q)$.}
\label{Fig5}
\end{figure}
In order to demonstrate that the Purcell factor fluctuations observed in the investigated prime arrays inherit the multifractal properties of the corresponding geometrical systems \cite{SgrignuoliMF}, we have analyzed the size-scaling of the gaps and also applied the MDFA method \cite{Ihlen,Kantelhardt} to the representative $P(\textbf{r}_e,\omega)$ (one-dimensional) profiles shown in panel (b) of Figs.\,(\ref{Fig2}) and (\ref{Fig3}). Figures\,(\ref{Fig4}) (a-c) show the scaling of the Purcell spectra with respect to the system size in a region surrounding the fundamental TM-band gap for the honeycomb, Eisenstein, and Gaussian structure, respectively. The appearance of more sub-gaps separated by a hierarchy of localized states is clearly manifested as a function of the number $N$ of nanocylinders in the systems. In panels (d-f) we quantify the system-size dependence of the TM fundamental gaps that are the most relevant ones for a pillar configuration \cite{Joannopoulos}. The mid-gap Purcell values for the honeycomb lattice, identified by the circular green markers, show an exponential scaling $e^{-\alpha L}$ with $\alpha=2.4$. This scaling is typical of periodic bandgap structures and was previously established in finite-size systems  \cite{AsatryanPRE,Leistikow}.
However, the corresponding analysis performed on the Eisenstein (red circular markers) and Gaussian (blue circular markers) arrays manifests instead a characteristic power-law scaling behavior $L^{-\beta}$ with $\beta$ equal to 15.8 and 14.5, respectively. 
Interestingly, we found that the values of the $\beta$ coefficients of both the Eisenstein and Gaussian arrays are different at their higher-frequency TM-gaps, while exactly the same values of $\alpha$ are found at the higher-frequency TM-gaps of the honeycomb lattice, as reported in the supplementary information. These results indicate a non-uniform power-law scaling of the LDOS of the Eisenstein and Gaussian arrays, which is linked to their multifractal nature. Moreover, these two different scaling laws reflect the different nature of the band-edge resonances of the investigated aperiodic structures. In fact, while band-edge resonances in periodic crystals are extended states with a spatial extension that scales linearly with the system size, the band-edge resonances of prime arrays are critical modes exhibiting local fluctuations and spatial oscillations over multiple-length scales \cite{SgrignuoliMF,WangPRB,MaciaBook,DalNegroBook,prado2021structural}. Moreover, self-similar critical modes decay weaker than exponentially, most likely with a power law \cite{Fujiwara,DalNegrocrystal2016,Macia}.   

In order to further characterize the strong multifractal scaling of the LDOS of prime arrays we apply the multifractal detrended fluctuation analysis (MDFA) \cite{Kantelhardt} to the simulated LDOS spectra using the numerical routines developed by Ihlen for the study of non-stationary complex signals \cite{Ihlen}.
The MDFA is a powerful characterization technique that generalizes the traditional detrended fluctuation analysis (DFA) \cite{peng1994mosaic} to non-stationary time series with multifractal properties by considering the scaling properties of their fluctuations with respect to local trends. We remind that in contrast to a homogeneous fractal (or monofractal), the scaling behavior around any point of a general multifractal measure $\mu$ is described by its local (i.e., position dependent) power law scaling $\mu({\vec {x}}+{\vec {a}})-\mu({\vec {x}})\sim a^{h({\vec {x}})}$ where the generalized Hurst exponent $h({\vec {x}})$, or singularity exponent, quantifies the strength of the singularity of the signal around that point. Moreover, the set of all the points that share the same singularity exponent is a fractal set characterized by a continuous distribution of fractal dimensions, i.e., by its singularity (or multifractal) spectrum \cite{falconer2004fractal}. 
The MDFA enables the accurate determination of the multifractal parameters of a signal, including its multifractal spectrum, by considering the local scaling of generalized fluctuations with respect to smooth trends obtained by piecewise sequences of locally approximating polynomial fits (typically obtained by straight-line fits), i.e., ${\displaystyle F_{q}(n)\propto n^{h(q)}}$ where $h(q)$ is the generalized Hurst exponent, or $q$-order singularity exponent \cite{Ihlen}. (Note that $h(q)$ reduces to the conventional Hurst exponent $H\in[0,1]$ for stationary signals).
Here the multifractal signal $x_{t}$ is regarded as a series of data points labelled by the integer parameter ${t}\in\mathbb{N}$ (corresponding to the discretized frequency in the LDOS case) and the generalized fluctuations $F_{q}(n)$ are taken as the $q$-order moments defined over $N$ intervals of size $n$ according to \cite{Kantelhardt,Ihlen}:
\begin{equation}\label{moment}
{\displaystyle F_{q}(n)=\left({\frac {1}{N}}\sum _{t=1}^{N}\left(X_{t}-Y_{t}\right)^{q}\right)^{1/q}}
\end{equation}
where:
\begin{equation}
X_t=\sum_{k=1}^t (x_k-\langle x\rangle)    
\end{equation}
Moreover, ${\displaystyle X_{t}}$ is subdivided into time windows of size ${\displaystyle n}$ samples and ${\displaystyle Y_{t}}$ indicates the piecewise sequence of approximating trends obtained by local least squares fits. Finally, ${\displaystyle \langle x\rangle }$ denotes the mean value of the time series corresponding to the analyzed signal. We note that the standard DFA approach can be obtained by setting $q=2$ in the eq.\,(\ref{moment}).

In Fig.\,(\ref{Fig5}) we report the most relevant quantities that are  used to characterise multifractal structures evaluated using the MDFA method for two representative LDOS spectra corresponding to the Eisenstein (a-c) and Gaussian arrays (d-f), respectively. In particular, the $q$-order Hurst exponents $H(q)$ shown in panels (a,d) are used to compute the $q$-order mass exponents $\tau(q)$ according to the relation \cite{Kantelhardt} $\tau(q)=qH(q)-1$. The nonlinearity of the mass exponents confirms the multifractal scaling nature of the analyzed LDOS signals \cite{nakayama2003fractal,stanley1988multifractal,hilborn2000chaos}. Moreover, the multifractal spectra of the LDOS, shown in panels (c,f), are derived from the mass exponent $\tau(q)$ via the Legendre transform \cite{Kantelhardt}:
\begin{equation}
D(q)={q}\tau^{\prime}(q)-\tau(q)  
\end{equation}
The broad and downward concavities of the multifractals spectra of the LDOS are typical of the strong multifractal behavior of Eisenstein and Gaussian primes \cite{SgrignuoliMF}. Since the  widths of the multifractal spectra are directly related to the degree of non-uniformity of the corresponding signals, our results demonstrate the stronger multifractality of the LDOS of Eisenstein arrays. 

Spectral fractality has profound implications on the relaxation dynamics of quantum sources \cite{akkermans2013spontaneous}. Here we address the decay rate of a two-level atom (TLA) with transition frequency $\omega_{if}$ embedded inside a structured vacuum with multifractal scaling properties. The time evolution of its excited state probability amplitude $C_2(t)$ obeys the general integro-differential equation \cite{vogel2006quantum}
\begin{equation}
\partial_t{C}_{2}(t)=\int_{0}^{t} \mathrm{~d} t^{\prime} K\left(t-t^{\prime}\right) C_{2}\left(t^{\prime}\right) \label{Eq:6}
\end{equation}
where the function $K(t)$ is a memory kernel that depends on the Green's tensor of the medium \cite{vogel2006quantum}.
In a multifractal structure $K(t)$ encodes the multi-scale fluctuations of the partial local density of states $\rho_p(\textbf{r}_e,\omega)$ \cite{Novotny}. In fact, within the rotating wave approximation \cite{Dung}, we can directly connect $K(t)$ with $\rho_p(\textbf{r}_e,\omega)$, resulting in the expression:
\begin{equation}
K(t)=-\frac{\omega_{if}|\textbf{p}|^{2}}{6\hbar\epsilon_0}\int_0^\infty d\omega e^{-i(\omega-\omega_{if})t}\rho_p(\textbf{r}_e,\omega) \label{Eq:8}
\end{equation}
where $\textbf{p}$ is the transition dipole moment of the TLA. Therefore, the usual Markovian approximation that is used in the Weisskopf-Wigner quantum theory of spontaneous decay fails in a multifractal vacuum. Moreover, the decay rate of the TLA can be significantly enhanced by  multifractal spectra. This can simply be appreciated by substituting eq.\,(\ref{Eq:8}) into eq.\,(\ref{Eq:6}), which yields the expression:
\begin{equation}
    \Gamma(t)=\frac{\omega_{if}|\textbf{p}|^{2}}{3\hbar\epsilon_0}\int_0^\infty d\omega e^{-\frac{i}{2}(\omega-\omega_{if})t}\frac{\sin(\omega-\omega_{if})t/2}{(\omega-\omega_{if})/2}\rho_p(\textbf{r}_e,\omega) \label{Eq:9}
\end{equation}
\begin{figure*}[t!]
\centering
\includegraphics[width=\linewidth]{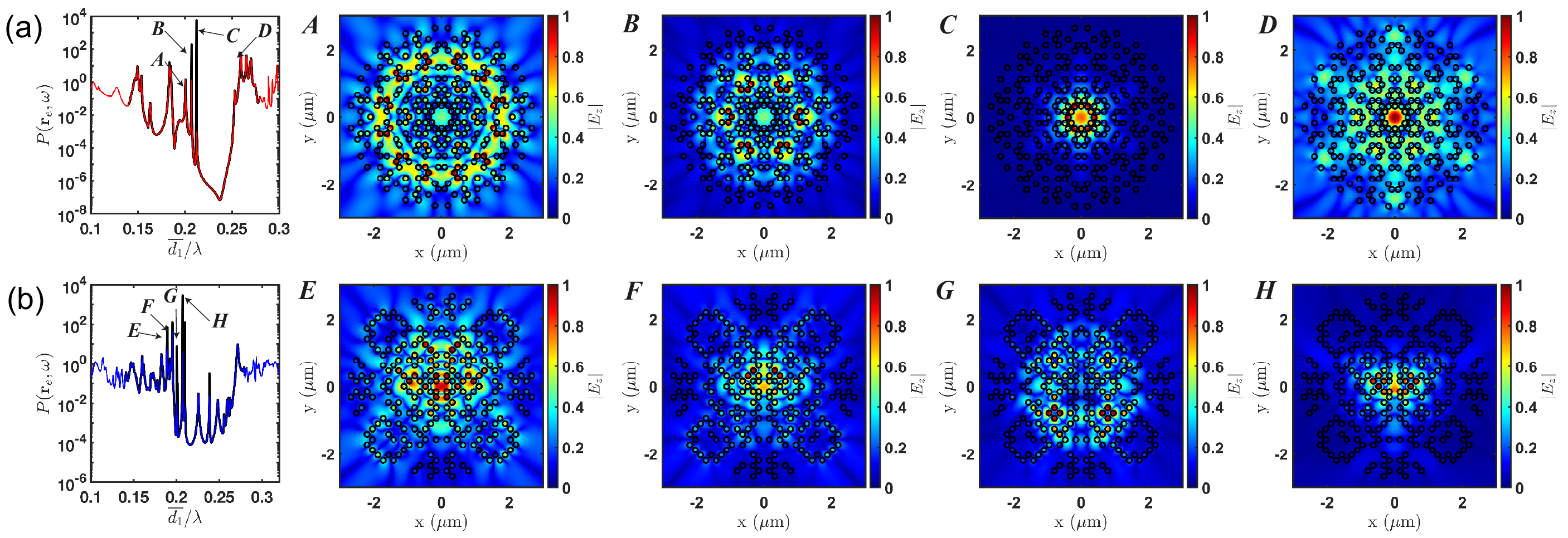}
\caption{Panel (a) and panel (b) show representative TM-polarized optical modes (i.e., $z$-component of the electric field) of the Eisenstein and the Gaussian prime configuration, respectively. The square root is used to enhance the contrast. Numbers are used to identify the spectral locations of the selected modes. The black curves in panels (a-b) are evaluated by increasing the spectral resolution up to $3\times10^{-6}\overline{d}_1/\lambda$ around the bandgap region. This high-resolution is needed to properly characterize the most localized modes near the band edges. The quality factors $Q$ of the selected modes in the Eisenstein configuration are equal to $Q_{A}=10^3$, $Q_{B}=7\times10^4$, $Q_{C}=2\times10^{12}$, and $Q_{D}=78$; while the Gaussian modes \textit{E}-\textit{H} have Q-factors equal to $2\times10^{3}$, $5\times10^{3}$ , $2\times10^{4}$, and $10^5$, respectively.} 
\label{Fig6}
\end{figure*}
where $\Gamma(t)={-[\partial t|{C}_{2}(t)|^{2}}]/{|{C}_{2}(t)|^{2}}$ is the decay rate of the excited state amplitude. Following Akkermans and Gurevich \cite{akkermans2013spontaneous}, we introduce the smooth spectral measure $\mu(\omega)$ associated to the integrated density of states (IDOS) and the density of modes $d\mu(\omega)=\rho_pd\omega$.
Expanding around a fixed point $\omega_{if}$, the multifractal measure is characterized by the local power-law scaling $\mu(\omega)\sim\epsilon^{d_{s}(\omega)}$ where $d_{s}(\omega)$ is the spectral dimension of the system around $\omega_{if}$ \cite{damanik2011spectral}. Considering the decay rate in the small frequency interval $\omega_{if}-\epsilon/2\le\omega\le\omega_{if}+\epsilon/2$, where $\epsilon\sim t^{-1}$, we obtain from eq.\,(\ref{Eq:9}) an estimate for the decay rate:
\begin{equation}
    \Gamma_{\omega_{if}}(t)=\frac{\omega_{if}|\textbf{p}|^{2}}{3\hbar\epsilon_0}t\int^{\omega_{if}+\epsilon/2}_{\omega_{if}-\epsilon/2}d\mu(\omega)\sim\phi(\omega_{if}) t^{1-d_{s}(\omega)} \label{Eq:11}
\end{equation}
where $\phi(\omega_{if})=\omega_{if}|\textbf{p}|^{2}/3\hbar\epsilon_0$ is a constant. This result agrees with what previously derived for fractal systems \cite{akkermans2013spontaneous}. In particular, a stretched exponential decay for the excited state probability is obtained as follows \cite{akkermans2013spontaneous}:
\begin{equation}
    |C_{2}(t)|^{2}=\exp\left[-\int_{0}^{t}\Gamma_{\omega_{if}}(t^{\prime})dt^{\prime}\right]=\exp\left[-\phi(\omega_{if})t^{\beta}\right]
\end{equation}
where $\beta=2-d_{s}(\omega)$.  

In this work, using the accurate Green's function method, we focus on the fundamental consequences of multifractal density of states on mode localization and two-photon emission processes, as discussed in the remaining sections.  
\section{Localized optical modes analysis}

In order to characterize the electromagnetic modes supported by the prime arrays we have solved the homogeneous T-matrix equation $\textbf{T} \textbf{b}=0$ \cite{Trojak1,Trojak2,DalNegroReview,TrevinoNano,Boriskina1}. The resonant modes of the structures have been obtained by numerically finding the complex $k$ values for which
the relation $\det[\textbf{T}(\tilde{k})]=0$ is satisfied \cite{Gagnon}. The complex eigenvalue $\tilde{k}$ has a physical interpretation in scattering theory, namely its real part is equal to the free-space wave number of the mode while its imaginary part describes its decay rate, which is inversely proportional to the spectral width of the mode. Modes with imaginary parts closer to zero are the most strongly localized and manifest  the highest quality factors. In this work we have evaluated the resonant optical modes by generating two-dimensional maps of $\det[\textbf{T}(k,\Im[k])]$ with a very large spectral resolution around the narrow LDOS peaks labelled (and highlighted in black) in Fig.\,(\ref{Fig6}). Specifically, we have used a resolution of $\Delta\Re{k}$ equal to $3\times10^{-4}\mu m^{-1}$ and a spacing in log-space of $\Delta[\log_{10}\Im(k)]$ equal to 0.07. The $T$-matrix was calculated by truncating the multipolar expansion order to $\ell_{max}=2$ (i.e., the field summations run from $-\ell_{max},\cdots,+\ell_{max}$, yielding a total of $2\ell_{max}+1=5$ multipolar orders). We have ensured that that this truncation accurately describes the optical properties of the investigated arrays for the considered size parameters and interparticle separations discussed in the paper \cite{Bohren}. 

Figure\,(\ref{Fig6}) shows the $z$-component of the electric field of the most localized band-edge resonant modes of the Eisenstein (a) and Gaussian (b) arrays. These modes correspond to the different Purcell enhancement peaks that are indicated by the letters $A-D$  and $E-H$ for the Eisenstein and Gaussian configurations, respectively. Similarly to the case of periodic photonic structures, the modes located on the low-energy band edges feature an electric field distribution that is largely concentrated inside the high-$\epsilon$ regions, while the ones closer to high-energy edges are mostly concentrated in the air-host regions \cite{Joannopoulos}. The quality factors $Q$, evaluated through the general expression $Q=|\Re(k)/2\Im(k)|$ \cite{Trojak1}, of the air modes confirm their leaky (radiative) nature with values as small as $Q_D=78$. In contrast, the lower-energy dielectric modes exhibit very large quality factors (values reported in the caption) up to $\sim{10^{12}}$, which are characteristic of strongly localized mode patterns. More generally, the optical resonances of Eisenstein and Gaussian photonic arrays show a large spectrum of complex spatial distributions that range from extended critical resonances, such as modes $(A,B,D)$ and $(E,F,G)$, to strongly localized ones confined at the center of the arrays (e.g., modes $C$ and $H$). To further demonstrate the localized nature of modes $C$ and $H$, we report in the supplemental document a characterization based on the mode spatial extent parameter. This study confirms that the modes $C$ and $H$ are the most localized ones in the Eisenstein and Gaussian configurations, respectively. The spatial distributions of additional electromagnetic modes supported by these structures are also reported in the supplemental document. The extended optical resonances of Eisenstein and Gaussian arrays are examples of critical resonances characterized by multifractal intensity fluctuations \cite{WangPRB,SgrignuoliMF}. The distinctive multi-scale and multi-frequency localization properties of critical resonances observed in these novel photonic systems provide opportunities to enhance light-matter interactions over broader spectral ranges compared to photonic crystal structures. These results motivate an alternative approach to enhance the generation efficiency of two-photon spontaneous emission processes, which we address in  the next section.

\section{Two-photon spontaneous emission rate}
In this section, we will discuss the relation between the TPSE rate and the modification of the spontaneous decay rate of an emitter due to its environment in the weak coupling regime (i.e., Purcell enhancement). Let us consider a quantum emitter located close to an arbitrary shaped two-dimensional nano-structures and study its decay from an initial state of energy $E_i=\hbar\omega_i$ to a final one $E_f=\hbar\omega_f$ via two-quantum processes assisted by intermediate states $\ket{I}$ of energy $\hbar\omega_n$. The Hamiltonian of the system "emitter+field" is given, within the rotating wave approximation, by $H=H_a+H_f+V$, where $H_a$ is the Hamiltonian of the isolated emitter, $H_f$ is the free-field Hamiltonian, and $V$ describes the emitter-field interaction, which has to be treated using perturbative methods \cite{Cohen}.  According to the Fermi's golden rule, the spontaneous emission rate $\Gamma$ between an initial state $\ket{i}$ and a final state $\ket{f}$ is given by \cite{Novotny}:
\begin{equation}\label{FermiGR}
\Gamma_{i\rightarrow f}=\frac{2\pi}{\hbar}|M_{fi}|^2\delta(E_f-E_i)
\end{equation}
Initially, the emitter is in the excited state and the field is in the vacuum (i.e, initial) state: the emitter-field initial state is $\ket{i}=\ket{e;0}$. Since we are interested in TPSE processes, the final emitter-field state is characterized by the emitter in a lower energy state and the field in a two-photon state, namely, $\ket{f}=\ket{g;1_\alpha,1_\alpha^\prime}$. The intermediate state $\ket{I}$ can be expressed as $\ket{n;1_\alpha}$ or $\ket{n,1_\alpha^\prime}$, where the index $n$ refers to the emitter states. Therefore, the transition matrix $M_{fi}$ in eq.\,(\ref{FermiGR}) can be written as \cite{MunizPRA}:
\begin{equation}\label{Mfi}
M_{if}=\sum_I\frac{\bra{f}V\ket{I}\bra{I}V\ket{i}}{\hbar\omega_f-\hbar\omega_i}
\end{equation}  
where $V=-\textbf{p}\cdot\textbf{E}$ \cite{Novotny}. Here, $\textbf{p}$ and $\textbf{E}$ are the dipole moment operator and the electric field operator. 
When the dominant transition wavelengths are much larger than the emitter's dimensions it is accurate to use the dipole approximation. Moreover, using the spectral representation of the Green's tensor $\mathbb{G}(\textbf{r},\textbf{r}^\prime;\omega)$ of the Helmholtz's equation \cite{Novotny}, we can express the TPSE rate $\Gamma^{(2)}(\textbf{r}_e)$ in the form \cite{MunizPRA, MunizPRL,Poddubny}:
\begin{eqnarray}\label{TPSE_final}
\Gamma^{(2)}(\textbf{r}_e)=\frac{\mu_0^2}{\pi\hbar^2}\int_0^{\omega_{if}}d\omega\sum_{i,l,j,k}\omega^2(\omega_{if}-\omega)^2\mathbb{D}_{ij}(\omega,\omega_{if}-\omega)\nonumber\\
\mathbb{D}^*_{lk}(\omega,\omega_{if}-\omega)\Im\mathbb{G}_{jk}(\textbf{r}_e,\textbf{r}_e;\omega_{if}-\omega)\Im\mathbb{G}_{il}(\textbf{r}_e,\textbf{r}_e;\omega)
\end{eqnarray}
where $\omega_{if}$ is the transition frequency between the initial and the final state, while the tensor quantity $\mathbb{D}$ is defined as \cite{MunizPRA,MunizPRL}:
\begin{equation}
\mathbb{D}(\omega_\alpha,\omega_{\alpha^\prime})=\sum_{n}\left[\frac{\textbf{p}_{in}\textbf{p}_{nf}}{\omega_{in}-\omega_{\alpha}}+\frac{\textbf{p}_{nf}\textbf{p}_{in}}{\omega_{in}-\omega_{\alpha^\prime}}\right]
\end{equation}
with $\textbf{p}_{nn^\prime}=\bra{n}\textbf{p}\ket{n^\prime}$, $\omega_{nn^\prime}=(E_n-E_{n^\prime})/\hbar$, and the summation extends over all the emitter's intermediate states (i.e., $n\neq i,f$). 

In free space, eq.\,(\ref{TPSE_final}) reduces to \cite{Craig}:
\begin{equation}
\Gamma_0^{(2)}=\int_0^{\omega_{if}}d\omega \gamma_0^{(2)}(\omega)
\end{equation}
where $\gamma_0^{(2)}$ is the free-space TPSE spectral density defined as follows \cite{MunizPRL,MunizPRA,Craig}:
\begin{equation}\label{TPSEvacuum}
\gamma_0^{(2)}(\omega)=\frac{\mu_0^2\omega^3}{36\pi^3\hbar^2c^2}(\omega_{if}-\omega)^2\abs{\mathbb{D}(\omega,\omega_{if}-\omega)}^2
\end{equation}
with $|\mathbb D(\omega,\omega_{if}-\omega)|^2=\mathbb D_{ij}(\omega,\omega_{if}-\omega)\mathbb D^*_{ij}(\omega,\omega_{if}-\omega)$. The term $\gamma_0(\omega)d\omega$ provides the number of emitted photons per unit of time (and unit volume) in the frequency interval $[\omega,\omega+d\omega]$. Due to the homogeneity of free space, $\Gamma_0^{(2)}$ is independent of the emitter's position. 
Crucially, the integrand of eq.\,(\ref{TPSE_final}), which is the TPSE spectral density $\gamma^{(2)}(\textbf{r}_e,\omega)$, can be expressed as a function of the structure-dependent Purcell enhancement yielding \cite{MunizPRL}:
\begin{equation}\label{TPSErate}
\gamma^{(2)}(\textbf{r}_e,\omega)=\gamma_0^{(2)}(\omega)\sum_{ij} t_{ij}(\omega) P_i(\textbf{r}_e,\omega)P_j(\textbf{r}_e,\omega_{if}-\omega)
\end{equation} 
where we used the eq.\,(\ref{TPSEvacuum}), the definition of the Purcell enhancement, and introduced the tensor $t_{ij}(\omega)=|\mathbb D_{ij}(\omega,\omega_{if}-\omega)|^2/|\mathbb D(\omega,\omega_{if}-\omega)|^2$, which depends only on the electronic structure of the emitter \cite{MunizPRL}. Here, $P_k(\textbf{r}_e,\omega)$ is the Purcell factor for a transition dipole moment oriented along the direction of the unit vector $\hat{\textbf{e}}_k$ with $k=x,y,z$. We recognize that in contrast to the one-photon SE, TPSE is a broadband phenomenon \cite{MunizPRA}. Another important feature of eq.\,(\ref{TPSErate}) is its symmetry with respect to $\omega_{if}/2$, namely, $\gamma^{(2)}(\textbf{r}_e,\omega)=\gamma^{(2)}(\textbf{r}_e,\omega_{if}-\omega)$, which is a direct consequence of the energy conservation since the sum of the frequencies of the two emitted photons must be equal to the emitter's transition frequency $\omega_{if}$. Equation\,(\ref{TPSErate}) provides the possibility to enhance the TPSE rate by properly engineered photonic environments with large Purcell enhancement values. 
\begin{figure*}[t!]
\centering
\includegraphics[width=14cm]{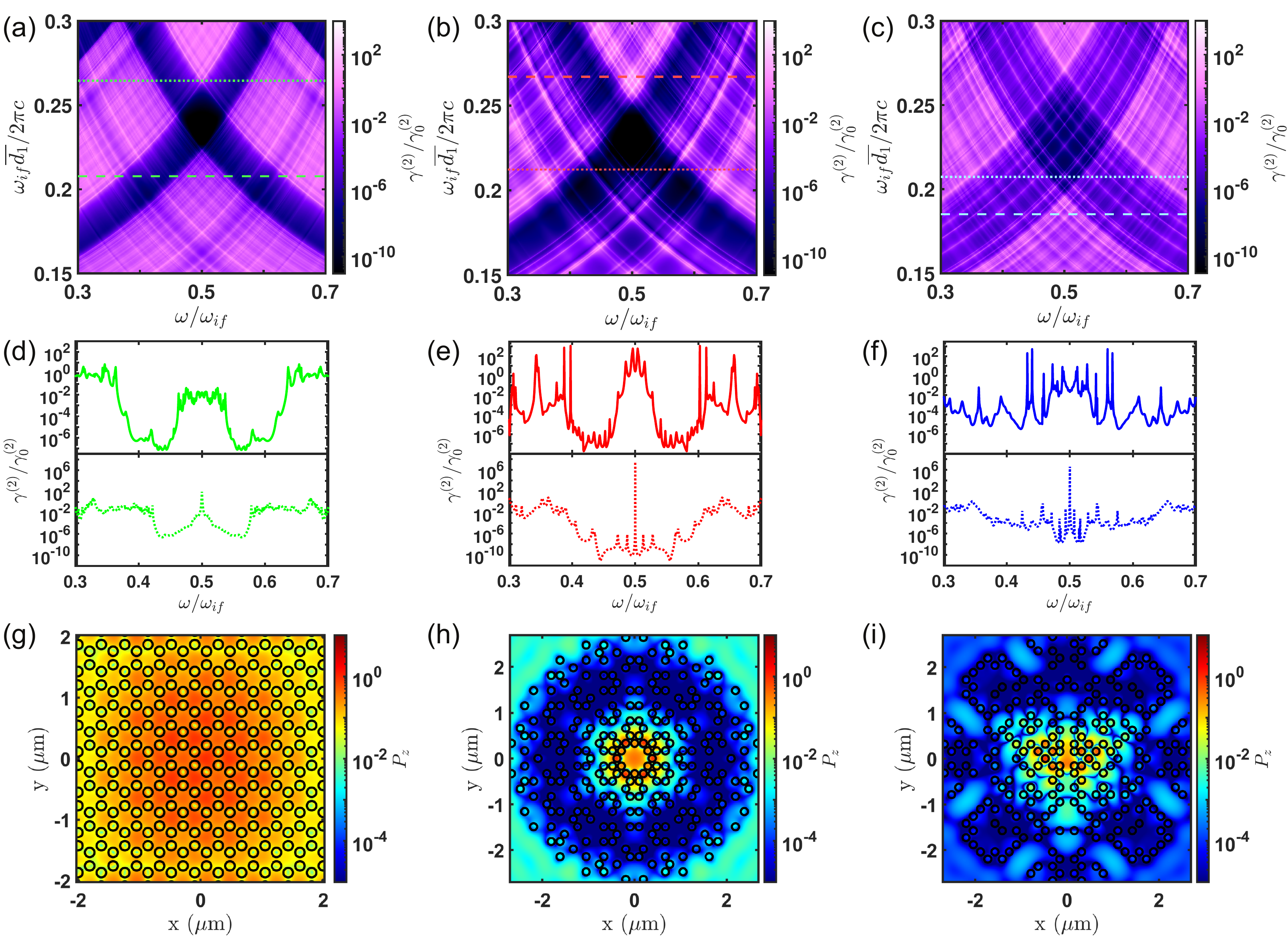}
\caption{ $\gamma^{(2)}/\gamma_0^{(2)}$ maps as a function of the dimensionless variables $\omega_{if}\overline{d}_1/2\pi c$ and $\omega/\omega_{if}$ when the nanocylinders are spatially arranged in the honeycomb lattice (a), Eisenstein (b) and Gaussian (c) prime arrays. These 2D-maps are evaluated by placing an emitter at the center of the investigated structures and are characterized by a resolution of $2.5\times 10^{-5}$ and $1.5\times10^{-6}$ for $\omega_{if}\overline{d}_1/2\pi c$ and $\omega/\omega_{if}$, respectively. Panels (d-f) show spectral TPSE profiles corresponding to the horizontal lines in panels (a-c). Green, red, and blue lines refer to the honeycomb, Eisenstein, and Gaussian structures. Panels (g-i) display the spatial distributions of the Purcell enhancement factors evaluated with a $10~nm$ spatial resolution and at the frequencies that maximize the TPSE spectral density [dashed lines in panel (a-c)] for honeycomb, Eisenstein, and Gaussian array.}
\label{Fig7}
\end{figure*}

Figure\,(\ref{Fig7}) (a-c) show high-resolution two-dimensional maps of $\gamma^{(2)}/\gamma_0^{(2)}$ as a function of the dimensionless variables $\omega_{if}\overline{d}_1/2\pi c$ and $\omega/\omega_{if}$. These maps are computed around the lowest energy band-edge region considering a two-level atom located at the center of approximately three hundred nanocylinders spatially arranged in a honeycomb lattice (a), Eisenstein (b), and Gaussian (c) prime arrays. The black regions identify the lowest TPSE rate values, reflecting the fundamental bandgaps of the investigated structures. The fragmentation of the Eisenstein and Gaussian principal bandgaps into a hierarchy of smaller sub-gaps is clearly visible also in panels (b) and (c). The multifractal nature of the band-edge regions of these structures is characterized by the presence of a large number of localized band-edge modes with significant Purcell enhancement across a large frequency spectrum, resulting in significantly larger TPSE rates than the investigated periodic system. These spectral enhancement maps strongly depend on the characteristics of the Purcell spectra of each investigated structure and are independent of the specific nature of the emitters. In Figs.\,(\ref{Fig7})(d-f), we show two representative TPSE spectra computed for the honeycomb lattice (panel d), Eisenstein (panel e), and Gaussian (panel f) arrays corresponding to the dotted and dashed lines that are visible in panels (a-c). These lines identify the structural configurations that yield the largest number of resonant $\gamma^{(2)}/\gamma_0^{(2)}$ peaks (top panels d-f) and the maximum TPSE enhancement values (lower panels d-f) that can be achieved in each investigated structure, respectively. Specifically, the top panels in Figs.\,(\ref{Fig7})\,(d-f) demonstrate that the prime arrays feature highly fragmented TPSE spectra with a large number of resonant peaks that show enhancement values larger than unity at different frequencies across their multifractal band-edge regions. In contrast, the honeycomb lattice  feature more regular TPSE spectra (i.e., continuous green lines) with values of $\gamma^{(2)}$ that are lower than the free-space value $\gamma_0^{(2)}$ except for the frequencies where band-edge modes are present. Moreover, due to the enhanced spatial localization of their critical modes, the values of $\gamma^{(2)}$ achieved by the Eisenstein (i.e., continuous red line) and Gaussian (i.e., continuous blue line) arrays significantly exceed the free-space value over the frequency range around $\omega=\omega_{if}/2$. Moreover, different side peaks appear at multiple frequencies in both panels \,(e) and \,(f), with $\gamma^{(2)}$ values that exceed their corresponding free-space ones by more than two orders of magnitude. This characteristic broadband TPSE enhancement behavior in prime arrays is a direct consequence of their multifractal density of states that supports a large number of highly resonant states inside narrow fractal sub-gaps, as discussed in the previous section (see also Fig.\,\ref{Fig6}). The bottom panels of Fig.\,(\ref{Fig7})\,(d-f) display the $\gamma^{(2)}/\gamma_0^{(2)}$ spectra with the maximum  enhancement for each investigated array. In particular, a strong peak at $\omega=\omega_{if}/2$ is clearly visible. However, an enhancement of more than $6$ orders of magnitude is observed for the Eisenstein (panel (e)) and Gaussian (panel (f))  structures, which is approximately $4$ orders of magnitude larger than what can be achieved by the  honeycomb photonic structure. This is a consequence of the more efficient wave localization mechanism in prime arrays where photons are resonantly scattered and localized at multiple-length scales \cite{SgrignuoliMF,SgrignuoliRW}. We also remark that our approach is different from what previously reported in the mid-IR range where localized plasmon- and phonon-polariton resonances can be efficiently excited in sub-wavelength polar materials or in metallic nanostructures \cite{Rivera,MunizPRL,MunizPRA}. 
Instead, here we demonstrate novel dielectric bandgap platforms for tailoring and enhancing TPSE using dielectric materials. 
In particular, due to their multi-scale structural complexity, Eisenstein and Gaussian prime arrays support multifractal spectra of highly-confined optical resonances that drive significant TPSE enhancement over a broad frequency range beyond what is possible using regular periodic structures.
The multi-frequency behavior of the TPSE enhancement of prime arrays compared to periodic systems with non-fractal mode density plays an important role in practical applications to quantum emitters that are always characterized by a finite emission linewidth. By comparing the spectral positions at which the TPSE is maximized with the peak values of the LDOS shown in Fig.\,(\ref{Fig6}), we realize that the values of $\omega_{if}\overline{d}_1/2\pi c$ for the Eisenstein [dashed red line in Fig.\,(\ref{Fig7}) (b)] and for the Gaussian array [dashed cyan line in Fig.\,(\ref{Fig7}) (c)] are driven by the optical modes labelled by the letters C and H in  Fig.\,(\ref{Fig6}). To establish that these highly confined critical modes indeed are responsible for the observed $\gamma^{(2)}/\gamma_0^{(2)}$ enhancement values, we evaluate the spatial profiles of the corresponding Purcell factors at these frequencies. The Purcell spatial maps are computed using a spatial grid of vertically-polarized electric dipoles with a resolution of $10~nm$ (i.e., $\sim 6.4\times 10^5$ dipoles are considered here). Convergence tests were performed with respect to the multipolar order expansion coefficient $\ell$, to ensure accuracy in the considered size parameters and interparticle separation ranges \cite{Bohren}. Panels (g-i) in Fig.\,(\ref{Fig7})  display the results of our analysis and show that in the honeycomb configuration the largest TPSE enhancement is produced by the excitation of extended Bloch band-edge modes, in the Eisenstein and Gaussian arrays the enhancement is driven by the most confined critical band-edge modes $C$ and $H$ shown in Fig.\,(\ref{Fig6}). Therefore, our results demonstrate that the excitation of critical modes at the multifractal band-edge states of Eisenstein and Gaussian photonic structures significantly enhances the generation rates of two-photon spontaneous emission. 

\section{Conclusions}
In conclusion, we have proposed a novel approach to enhance the generation of two-photon spontaneous emission processes by leveraging the multifractal structural and spectral properties of the distributions of Eisenstein and Gaussian primes. Using the rigorous Mie-Lorenz multipolar theory, we have analyzed their Purcell spectra and demonstrated that the prime arrays support  complete photonic bandgaps despite their lack of periodicity. In addition, we have established the multifractal nature of these bandgaps with a characteristic size-scaling behavior and fragmentation of spectral gap regions into a hierarchy of smaller sub-gaps separated by many highly localized modes. The multifractal spectra of the local density of states of Eisenstein and Gaussian structures have been characterized using multifractal detrended fluctuation analysis and their fundamental implications on the relaxation dynamics of quantum sources has been discussed. Moreover, using a general Green's tensor approach we systematically study the enhancement of the two-photon spontaneous emission rates and demonstrate that Eisenstein and Gaussian prime arrays give rise to broadband TPSE enhancement spectra compared to what is possible in periodic photonic structures of comparable size. Finally, we demonstrate that highly-localized critical modes across multifractal band-edge states of Eisenstein and Gaussian photonic structures can drive TPSE rates up to $6$ orders of magnitudes larger than what possible in free-space and $4$ orders of magnitudes larger than the values obtained for the honeycomb reference structure of comparable size.
Our findings establish the relevance of engineered aperiodic photonic designs based on the Eisenstein and Gaussian prime distributions for the efficient generation of multi-frequency two-photon processes in dielectric photonic devices. Besides defining a novel approach for enhanced quantum two-photon emitters, aperiodic bandgap structures with multifractal mode density may provide access to novel electromagnetic phenomena in structured environments with multifractal geometry for quantum nanophotonics applications.
\begin{acknowledgments}
L.D.N. acknowledges the support from the Army Research Laboratory under Cooperative Agreement Number W911NF-12-2-0023.
\end{acknowledgments}
\bibliographystyle{apsrev4-1}
%

\section{Multifractal scaling of Purcell factor: highest-frequency gaps}
\begin{figure*}[b!]
\centering
\includegraphics[width=13cm]{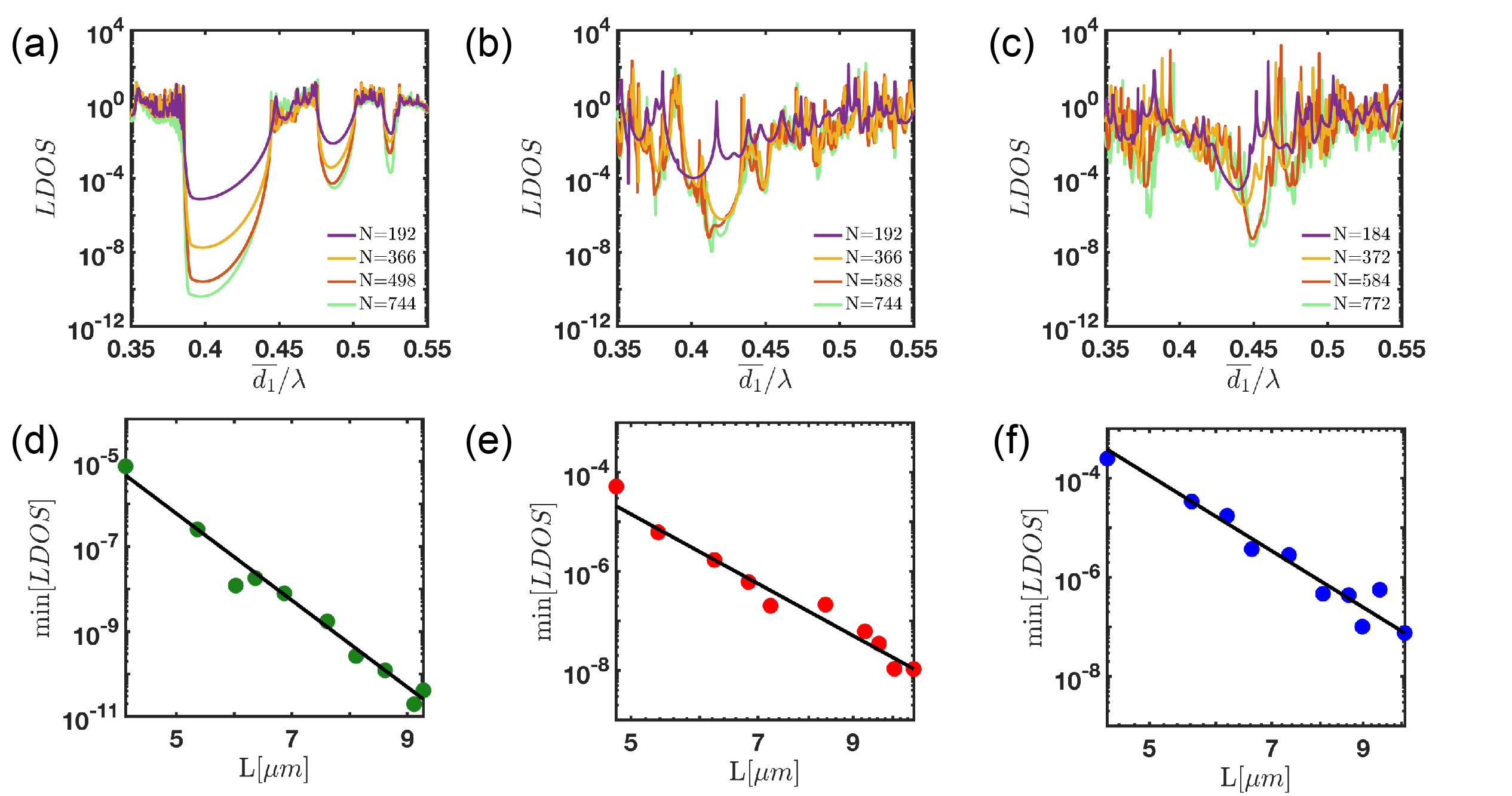}
\caption{Panels (a), (b), and (c) show the LDOS as a function of $\overline{d_1}/\lambda$ around the highest energy band gap region for a different number of dielectric nanocylinders, as specified in the corresponding legends, arranged in a honeycomb lattice, Eisenstein and Gaussian configuration, respectively. These LDOS profiles correspond to $r/\overline{d_1}=0.35$ and are evaluated by probing the different structures with a vertical line source located at the center. Panels (d-f) display the scaling of the minimum value of the LDOS inside the bandgap as a function of the system sizes. While the honeycomb lattice show an exponential scaling law $e^{\alpha L}$ with $\alpha=2.4$, the Eisenstein and Gaussian configuration are characterized by a power law decay $L^\beta$ with the scaling slope $\beta$ equal to 9.6 and 10.4, respectively.}
\label{FigS1}
\end{figure*}

Figure\,(\ref{FigS1}) shows the scaling of the Purcell factor with respect to the system size in a region surrounding the highest-frequency TM-bandgap of the honeycomb, Eisenstein, and Gaussian structure. Similarly to what observed for the TM-fundamental gaps, the scaling of the higher-frequency gaps are characterized by the appearance of more sub-gaps separated by a hierarchy of localized states both in the Eisenstein [see panel (b)] and Gaussian [see panel (c)] array. Moreover, panels (d), (e), and (f) quantify the scaling of the highest-frequency TM gaps for honeycomb, Eisenstein, and Gaussian, respectively. The mid-gap Purcell values of the honeycomb lattice feature an exponential scaling $e^{-\alpha L}$ with $\alpha=2.4$, which is the same as for the fundamental TM-gap (see Fig.4 of the main manuscript and relative discussion). This scaling is typical of periodic bandgap structures and was previously established in finite-size systems  \cite{AsatryanPRE,Hasan,Leistikow}. On the other hand, panels (e) and (f) display a power-law scaling $L^{-\beta}$ characterized by  different $\beta$ values for different gaps. This non-uniform power-law scaling is a consequence of the multifractal nature of prime arrays.

\begin{figure*}[b!]
\centering
\includegraphics[width=13cm]{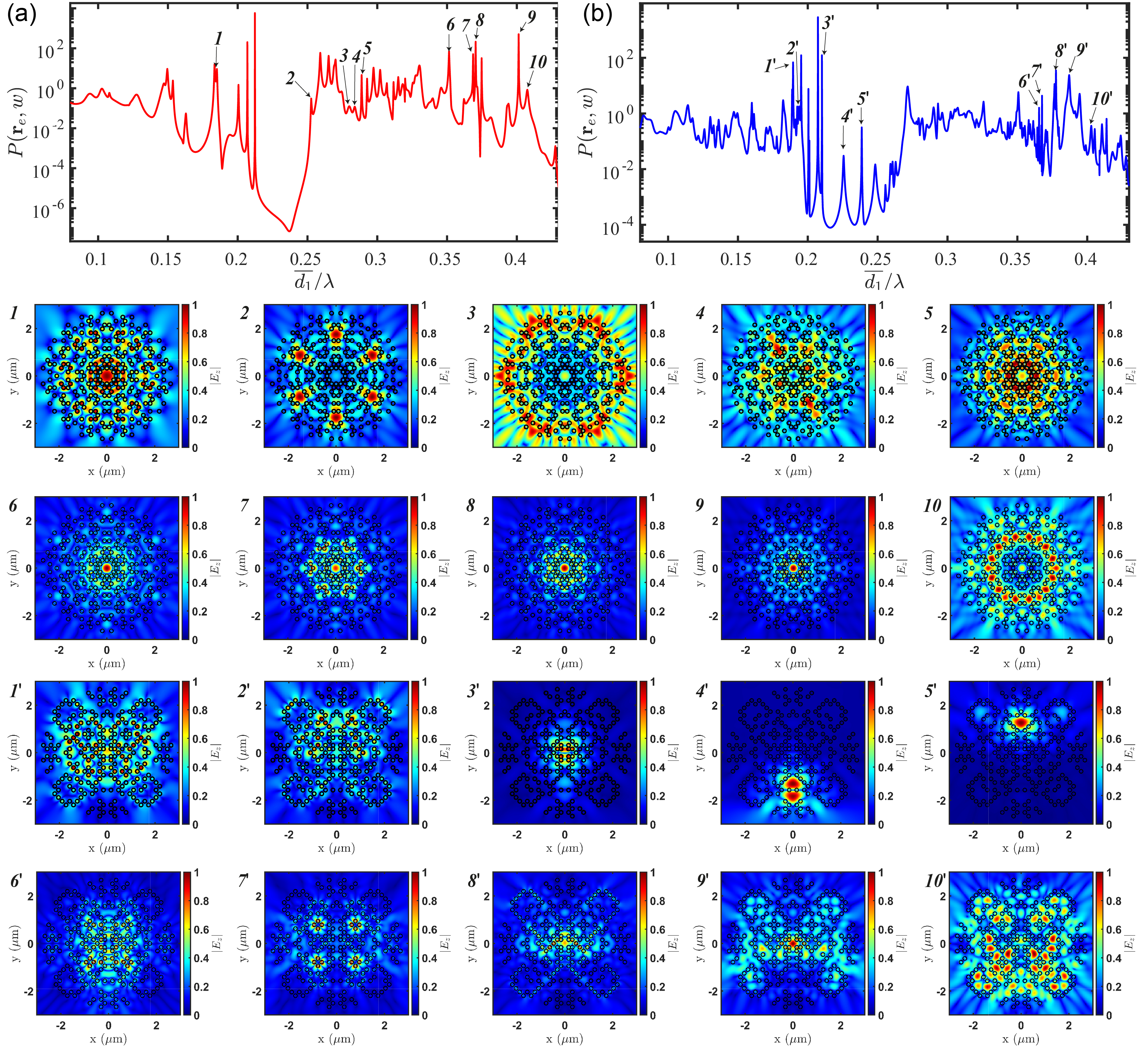}
\caption{Spatial distributions of TM-polarized optical mode (i.e., $z$-component of the electric field) surrounding the first two bandgaps of the Eisenstein (a) and Gaussian (b) array. The square root is used to enhance the contrast. Numbers are used to identify the spectral location of the selected modes evaluated with a spatial resolution of $12\,nm$. The resolution of the Purcell spectrum is $1.5\times10^{-6}\overline{d}_1/\lambda$. This high-resolution is needed to properly characterize the most localized modes near the investigated structures' band edges.}
\label{FigS2}
\end{figure*}

\section{Optical modes of prime arrays}\label{Modes}
Figures\,(\ref{FigS2}) and (\ref{FigS2}) display the spatial distributions of additional scattering resonances surrounding the first two bandgaps of the Eisenstein and Gaussian array, respectively. These scattering resonances are evaluated by using the same approach and the same parameters discussed in the main manuscript (see Fig.6 and the corresponding discussion for more details). The optical resonances of Eisenstein and Gaussian photonic arrays show a large spectrum of complex spatial distributions that range from extended critical modes to strongly localized resonances \cite{SgrignuoliMF}.

\section{Mode spatial extent of prime arrays optical modes}
In order to characterize the spatial extension of the optical modes discussed in the main manuscript and in the section\,\ref{Modes}, we have evaluated the mode spatial extent (MSE) parameter defined as \cite{MIRLIN2000259,SgrignuoliACS}:
\begin{equation}
MSE=\frac{\int|{E_z}(\mathbf{r})|^{4} \mathrm{~d}^{2} r}{\left(\int|{E_z}(\mathbf{r})|^{2} \mathrm{~d}^{2} r\right)^{2}}
\end{equation}
where the integral is performed in the $xy$ plane and $E_z$ denotes the electric field z-component (along the nanocylinder axis). The MSE parameter allows us to quantify the degree of localization of any optical modes \cite{MIRLIN2000259,skipetrov2014absence}. Table\,\ref{tbl:MSE} summarizes the results of this study applied to the modes presented in the main manuscript, while Table\,\ref{tbl:MSE2} shows the MSE of the resonances reported in Fig.\,\ref{FigS2}. This analysis displays that the modes $C$ and the mode $H$ are the most localized resonances around the lowest energy band gap region of Eisenstein and Gaussian configuration, respectively. This characterization corroborates our previous results based on the analysis of the quality factors related to these two modes presented in the main manuscript. Consistently, Table\,\ref{tbl:MSE2} shows that the MSE is lower for those modes that are spatially confined.          

\begin{table}[h]
  \centering
  \caption{Mode spatial extent for optical modes shown in Fig. 6}  
  \label{tbl:MSE}
  \begin{tabular}{ccccccccc}
    \hline
    Mode  & $A$ & $B$ & $C$ & $D$ & $E$ & $F$ & $G$ & $H$\\
    \hline
    MSE   & 3.99 & 1.50 & 0.58 & 2.97& 2.30 & 0.99 & 1.01 & 0.59\\
    \hline
  \end{tabular}  
\end{table}

\begin{table}[h]
  \centering
  \caption{Mode spatial extent for optical modes shown in Fig. \ref{FigS2}}  
  \label{tbl:MSE2}
  \begin{tabular}{ccccccccccc}
    \hline
    Mode  & $1$ & $2$ & $3$ & $4$ & $5$ & $6$ & $7$ & $8$ & $9$ & $10$\\
    \hline
    MSE   & 5.38 & 2.54 & 12.7 & 6.58 & 5.4 & 0.98 & 1.29 & 0.86 & 0.49 & 3.94\\
    \hline
    Mode  & $1'$ & $2'$ & $3'$ & $4'$ & $5'$ & $6'$ & $7'$ & $8'$ & $9'$ & $10'$\\
    \hline
    MSE & 3.64 & 2.56  & 0.52 & 0.62 & 0.35 & 1.54 & 0.69 & 0.75 & 1.87 & 4.42\\
    \hline
  \end{tabular}  
\end{table}

\end{document}